\documentclass[aps,prl,twocolumn,superscriptaddress,showpacs]{revtex4}

\usepackage{graphicx}
\usepackage{latexsym}
\usepackage{amssymb}
\usepackage{amsmath}

\begin{document}

\title{Dephasing Time of Two-Dimensional Holes in GaAs Open Quantum Dots}

\author{S. Faniel}
\email[]{faniel@pcpm.ucl.ac.be}
\affiliation{Cermin, PCPM and DICE Labs, Universit\'e Catholique de Louvain, 1348 Louvain-la-Neuve, Belgium}

\author{B. Hackens}
\affiliation{Cermin, PCPM and DICE Labs, Universit\'e Catholique de Louvain, 1348 Louvain-la-Neuve, Belgium}

\author{A. Vlad}
\affiliation{Cermin, PCPM and DICE Labs, Universit\'e Catholique de Louvain, 1348 Louvain-la-Neuve, Belgium}

\author{L. Moldovan}
\affiliation{Cermin, PCPM and DICE Labs, Universit\'e Catholique de Louvain, 1348 Louvain-la-Neuve, Belgium}

\author{C. Gustin}
\affiliation{Cermin, PCPM and DICE Labs, Universit\'e Catholique de Louvain, 1348 Louvain-la-Neuve, Belgium}

\author{B. Habib}
\affiliation{Department of Electrical Engineering, Princeton University, Princeton, New Jersey 08544, USA}

\author{S. Melinte}
\affiliation{Cermin, PCPM and DICE Labs, Universit\'e Catholique de Louvain, 1348 Louvain-la-Neuve, Belgium}

\author{M. Shayegan}
\affiliation{Department of Electrical Engineering, Princeton University, Princeton, New Jersey 08544, USA}

\author{V. Bayot}
\affiliation{Cermin, PCPM and DICE Labs, Universit\'e Catholique de Louvain, 1348 Louvain-la-Neuve, Belgium}

\date{\today}

\begin{abstract}
We report magnetotransport measurements of two-dimensional holes in open quantum dots, patterned either as a single-dot or an array of dots, on a GaAs quantum well. For temperatures $T$ below 500 mK, we observe signatures of coherent transport, namely, conductance fluctuations and weak antilocalization. From these effects, the hole dephasing time $\tau_\phi$ is extracted using the random matrix theory. While $\tau_\phi$ shows a $T$-dependence that lies between $T^{-1}$ and $T^{-2}$, similar to that reported for electrons, its value is found to be approximately one order of magnitude smaller. 
\end{abstract}

\pacs{73.21.La,73.23.Ad,03.65.Yz,85.35.Ds}

\maketitle
Spintronic devices~\cite{Spintronic}, which aim to manipulate carriers' spin, are promising candidates to replace classical electronic devices and have been intensively studied during the last decade. Within this context, characteristic transport times such as the dephasing time $\tau_\phi$ and the spin relaxation time are of paramount importance. Most investigations have focused on these characteristic times for electrons in two-dimensional systems, quantum wires, and quantum dots~\cite{Bird} for different materials including GaAs~\cite{dephasingGaAs,dephasingGaAs2} and InGaAs~\cite{dephasingInGaAs,BenPRL}. Up to now, only few studies have been carried out on p-type nanostructures~\cite{Holenano} because their experimental study is made difficult by the small amplitude of the holes' quantum interference effects. However, holes are good candidates for the emerging fields of spintronics and quantum computing. In particular, the use of holes has been suggested in quantum computing implementations because their spin should not couple to the nuclei~\cite{Loss}. Moreover, in GaAs systems, holes are subject to a stronger spin-orbit coupling than electrons~\cite{Eisenstein,Lu,Winkler}, making them attractive for spintronic applications. 

Here we report measurements of $\tau_\phi$ for holes in open quantum dots. Magnetotransport measurements display conductance fluctuations and weak antilocalization when the temperature $T$ is lowered below $\sim 500 \ \rm mK$. From these data, we determine $\tau_\phi$ using the random matrix theory. The temperature depencence of the measured $\tau_\phi$ lies between a $T^{-1}$ and $T^{-2}$ behavior, similar to that determined for electrons. Remarkably, its absolute value is approximately one order of magnitude smaller.

\begin{figure}[htbp]
  \includegraphics[width=1\columnwidth]{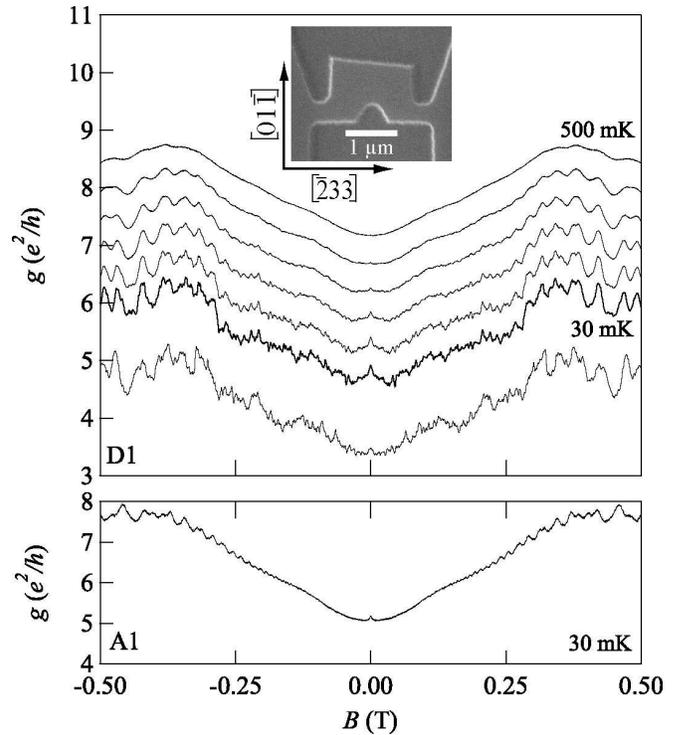}
  \caption{\label{figure1} Upper panel: magnetoconductance of D1 for $p = 2.3 \times 10^{15} \ \rm m^{-2}$ at $T =$ 30 (bold trace), 70, 130, 200, 300 and 500 mK. For clarity, traces are shifted upwards by $0.5 \ e^2/h$. A magnetoconductance trace measured in a second cooldown is also shown and is shifted downwards by $1.5 \ e^2/h$. Inset: SEM picture of D1. Lower panel: magnetoconductance of A1 for $p = 2.3 \times 10^{15} \ \rm m^{-2}$ at 30 mK. }
\end{figure}

The samples were fabricated from a p-type GaAs quantum well grown by molecular beam epitaxy on a $(311)A$ wafer. The two-dimensional hole system (2DHS) has a density $p = 2.3 \times 10^{15} \ \rm m^{-2}$ and a low-$T$ mobility of $35 \ \rm m^2/Vs$~\cite{footnote3}, equivalent to a mean free path of $2.7 \ \rm \mu m$. The 2DHS was contacted with Be/Au ohmic contacts. Two different dots with similar shapes but different areas ($1.4 \ \rm \mu m^2$ - D1 and $4.5 \ \rm \mu m^2$ - D2)~\cite{footnote2} were patterned using electron beam lithography and wet etching (inset to Fig.~\ref{figure1}). A back-gate and a Ti/Pt top-gate controlled the hole density, the shape of the vertical confining potential, and to some degree, the dots' openings. The measurements were performed down to 30 mK in a dilution refrigerator with the magnetic field $B$ applied perpendicular to the plane of the 2DHS. The conductance of the dots was measured using a standard lock-in technique at a frequency of 15 Hz, with a current of 1 nA. 

We first briefly comment on GaAs 2DHSs. In such systems, the spin-orbit interaction is strong and leads to a splitting of the valence band into heavy holes (spin = $\pm \ \frac{3}{2}$) and light holes (spin = $\pm \ \frac{1}{2}$). In the quantum well used to fabricate our samples, only the heavy hole subband is populated. Moreover, the spin-orbit interaction gives rise to a zero magnetic field spin-splitting. The magnitude of this spin splitting can be probed by Shubnikov - de Haas (SdH) measurements~\cite{Lu}. In this work, we investigate two different configurations: $p = 2.3 \times 10^{15} \ \rm m^{-2}$ with an asymmetric confining potential (the frequencies measured in the SdH oscillations are 3.7 and 5.3 T), and $p = 1.7 \times 10^{15} \ \rm m^{-2}$ where the quantum well is made symmetric by means of the gate voltages (the two frequencies then merge to the same value at 3.4 T). Note that in the latter configuration, even though the confining potential is symmetric and only one frequency is observed in the SdH data, the zero magnetic field spin-splitting is still present~\cite{PRLWinkler}. 

We first present the magnetotransport measurements performed on the open quantum dots. The conductance $g$ of D1 at $p = 2.3 \times 10^{15} \ \rm m^{-2}$ is plotted in the upper panel of Fig.~\ref{figure1} as a function of $B$ for various temperatures. At the lowest temperatures, we observe reproducible magnetoconductance fluctuations (MCFs), symmetric with respect to $B=0 \ \rm T$, that can be attributed to quantum interferences of holes inside the dot. When $T$ is increased, these MCFs are strongly reduced in amplitude and disappear for $T \gtrsim 500 \ \rm mK$. At these high $T$'s, only the slowly varying background remains, caused by ballistic effects in the cavity and the reduction of backscattering at the quantum point contacts~\cite{VanHouten}. From the mean conductance, we deduce that 5 to 6 modes are populated in each quantum point contact. We also note that for $B>0.25 \ \rm T$, SdH oscillations are visible on the dot's magnetoconductance traces. Similar data were obtained at $p = 1.7 \times 10^{15} \ \rm m^{-2}$ with two modes in each quantum point contact and also for D2 at $p = 2.3 \times 10^{15} \ \rm m^{-2}$ (not shown). Around $B=0 \ \rm T$, a sharp peak is observed in the conductance of D1 at low $T$ (Fig.~\ref{figure1}, upper panel). This peak is reminiscent of the weak antilocalization (WAL) correction to the conductivity. However, the superposition of the MCFs, which are comparable in magnitude, prevents any quantitative analysis of the WAL effect. This is clearly evidenced by a comparison of the WAL peak for two different cooldowns (Fig.~\ref{figure1}, upper panel). 

\begin{figure}[htbp]
  \includegraphics[width=1\columnwidth]{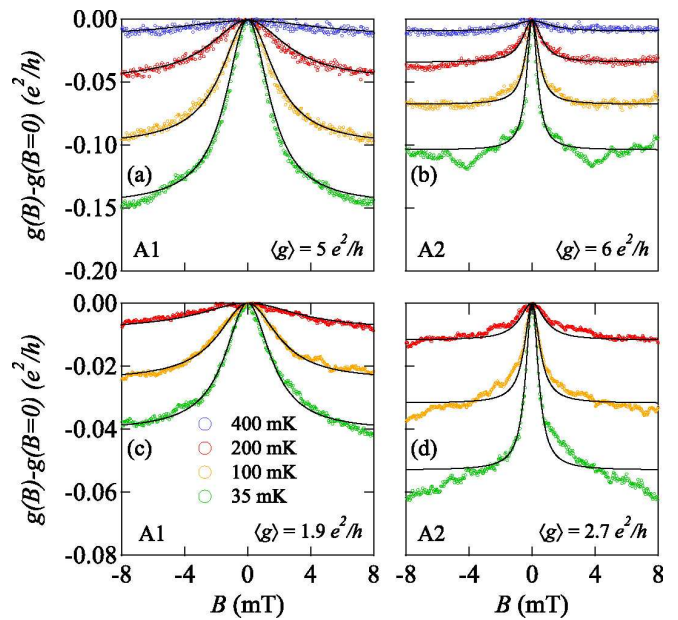}
  \caption{\label{figure2} (Color online) $g(B)-g(B=0)$ as a function of $B$ at indicated temperatures for (a) A1 and (b) A2 at $p = 2.3 \times 10^{15} \ \rm m^{-2}$ as well as (c) A1 and (d) A2 at $p = 1.7 \times 10^{15} \ \rm m^{-2}$. Solid lines show the fits of the WAL peak using the RMT. The mean conductance $\langle g \rangle$ at $B=0$ is given for each case.}
\end{figure}

In order to average out the MCFs and access the WAL correction to the conductivity~\cite{dephasingGaAs2,Antiloc}, we fabricated two additional samples made of arrays of 10x10 dots, spaced by $\rm 10 \ \mu m$, with cavities similar to D1 and D2. These samples are denoted as A1 and A2 for the smaller and the larger dots, respectively. As expected, the conductance of the arrays is made of a slowly varying background, similar to that of the single dots, but without MCFs (lower panel of Fig.~\ref{figure1}). For $T \lesssim 500 \ \rm mK$, a peak associated with WAL is observed in the magnetoconductance around $B=0 \ \rm T$ for both samples and for both investigated configurations (Fig.~\ref{figure2}). Because the holes' trajectories enclose a smaller magnetic flux in a dot with a smaller area, the width of the WAL peak is found to be larger in the case of A1 than A2. Note that our hole WAL peaks spread over a $B$ range approximatively $4$ times larger than in electron quantum dots with comparable areas~\cite{zumbuhl,Antiloc}. 

\begin{figure}[htbp]
  \includegraphics[width=1\columnwidth]{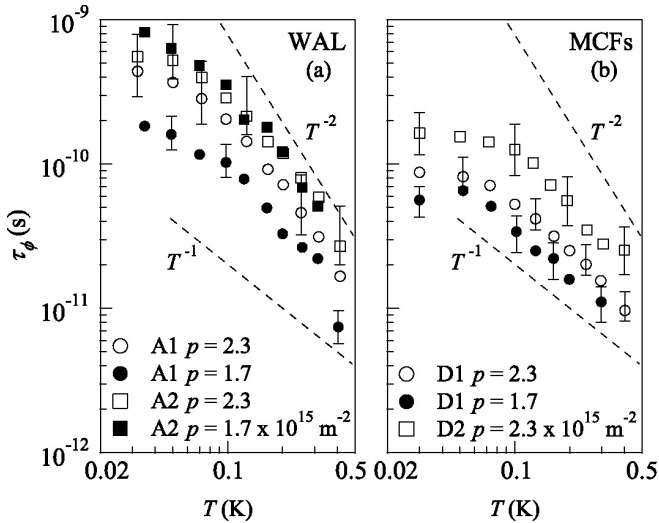}
  \caption{\label{figure4}Dephasing time $\tau_\phi$ of holes as a function of $T$. Left panel: $\tau_\phi$ extracted from the WAL correction to the conductivity in samples A1 and A2. Right panel: $\tau_\phi$ calculated from the variance of the MCFs in samples D1 and D2. The dotted lines indicate various $T$ dependencies. Error bars are determined from uncertainties in the number of modes in the quantum point contacts and in the dots' areas.}
\end{figure}

We now come to a more quantitative analysis of our data. Generally, electron transport in open quantum dots is described using the random matrix theory (RMT). To our knowledge, there has been no theoretical attempts to study the coherent transport of holes in such nanostructures. In particular, an appropriate model for magnetotransport in GaAs hole quantum dots, in the spirit of the theory developed for WAL in 2DHSs~\cite{HoleWAL2D}, would give access to both $\tau_{\phi}$ and the spin relaxation time. Although the RMT doesn't account for the complex band structure of GaAs hole systems under study, it has been recently extended in order to take the spin-orbit interaction into account~\cite{RMT1,RMT2}. We use this framework to analyze our data and start with the study of the WAL peak that provides information on both dephasing and spin-orbit interaction in the quantum dots. We fit the WAL peak using Eq. (13) of Ref.~\cite{RMT1} with  $m^{*} = 0.38 \ m_e$, where $m_e$ is the free electron mass. The fits are shown in Fig.~\ref{figure2} as solid curves. The three parameters of this model are $\tau_{\phi}$, $\tau_{so}$ and $c$, where $\tau_{so}$ is the spin-orbit scattering time and $c$ is a geometrical factor. For each sample and configuration, $c$ is determined from the fit to the lowest temperature traces. We obtain values ranging from $0.03$ to $0.06$. The fits indicate that $\tau_{so}$ is too small ($\lesssim \ 10^{-11} \ \rm s$) to be efficiently probed by the WAL in our samples and further confirm that the quantum dots are in a strong spin-orbit coupling regime. We therefore extract the hole $\tau_{\phi}$ in the quantum dots by setting $\tau_{so}=0$ in the fits. The $T$-dependence of $\tau_\phi$ in samples A1 and A2 is plotted in Fig.~\ref{figure4}(a) and is discussed below. 

\begin{figure}[htbp]
  \includegraphics[width=1\columnwidth]{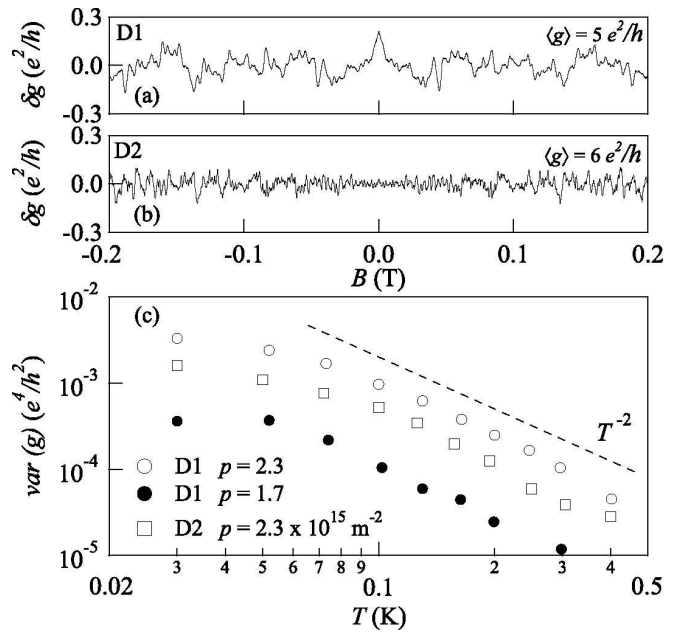}
  \caption{\label{figure3} Magnified view of the MCFs of (a) D1 and (b) D2 at $p = 2.3 \times 10^{15} \ \rm m^{-2}$ and $T = 30 \ \rm mK$ after background subtraction (see text). The mean conductance $\langle g \rangle$ is calculated in the range $-0.2 < B < 0.2 \ \rm T$. (c) Variance of the MCFs as a function of $T$ for D1 and D2. The dotted line indicates a $T^{-2}$ dependence.}
\end{figure}

The analysis of the MCFs also gives a measure of the carrier dephasing time. The RMT allows us to extract $\tau_\phi$ from the variance of the dot's conductance, $var(g)$, according to~\cite{RMT2}:
\begin{equation} 
var(g) = \int_{0}^{\infty} \int_{0}^{\infty} f'(E) f'(E') cov(E,E') dE dE' \ ,     
\label{var}
\end{equation}
\noindent \noindent where $E$ and $E'$ are energies, $f'(E)$ is the derivative of the Fermi function, and $cov(E,E')$ is the conductance correlator, given by Eq. (29) of Ref.~\cite{RMT2}. Before calculating $var(g)$, we isolate the MCFs from the background by applying a high-pass filter to the traces~\cite{footnote4}. Filtered traces are shown in Fig.~\ref{figure3} for D1 and D2 at $p = 2.3 \times 10^{15} \ \rm m^{-2}$ and $T=30 \ \rm mK$. Once the background is removed, we calculate the variance of the MCFs in the range $0.04 < B < 0.2 \ \rm T$. The MCFs variance is plotted in Fig.~\ref{figure3}(c) as a function of $T$ for D1 at $p = 2.3 \times 10^{15} \ \rm m^{-2}$ and $p = 1.7 \times 10^{15} \ \rm m^{-2}$ and for D2 at $p = 2.3 \times 10^{15} \ \rm m^{-2}$. In all cases, $var(g)$ exhibits a $T^{-2}$ behavior for $T \gtrsim 70 \ \rm mK$, and tends to saturate at lower $T$. Once the MCFs variance is extracted, we calculate the hole $\tau_\phi$ using Eq. (\ref{var}). As established by the analysis of WAL, we set $\tau_{so}=0$ in the calculation. The value of $\tau_\phi$ deduced from the MCFs is shown in Fig.~\ref{figure4}(b) as a function of $T$.

The hole dephasing time extracted from our data is shown in Fig.~\ref{figure4}. In Fig.~\ref{figure4}(a) we show $\tau_\phi$ extracted from the fits to the WAL peak of samples A1 and A2. Figure~\ref{figure4}(b) shows $\tau_\phi$ deduced from the variance of samples D1 and D2. Below $500 \ \rm mK$, $\tau_\phi$ exhibits a $T$-dependence that lies between $T^{-1}$ and $T^{-2}$ for all cases. While values of $\tau_\phi$ obtained from the MCFs analysis are slightly smaller and tend to saturate at low $T$ ($\lesssim 70 \ \rm mK$), results obtained from these two different methods are qualitatively in good agreement. 

The evolution of $\tau_\phi$ with $T$ is very similar to that reported for electrons in GaAs open quantum dots~\cite{Bird,dephasingGaAs,dephasingGaAs2}. However, the absolute value of $\tau_\phi$ is approximately one order of magnitude smaller in the case of holes. Generally, the expected mechanisms leading to dephasing at low $T$ in quantum dots are carrier-carrier scattering~\cite{dephasingGaAs2} as well as geometrically related mechanims such as dwell-time limiting effect~\cite{BenPRL} and environmental coupling~\cite{Bird1}. We recall that at temperatures where $\tau_\phi$ is not saturated, dephasing in electron quantum dots is well described by the Fermi liquid theory for 2D disordered systems~\cite{dephasingGaAs2,dephasingInGaAs}. Within the framework of this theory, $\tau_\phi$ resulting from carrier-carrier scattering~\cite{tauphi2D} is expected to decrease by a factor of $\sim 3$ for holes with respect to electrons because of their smaller Fermi energy and the lower mobility of the 2DHS. This reduction factor does not explain the small value of $\tau_\phi$ extracted for our quantum dots. However, Fermi liquid theory has been formulated for electrons and might not be directly applicable to 2D holes. Moreover, in 2DHSs other mechanisms like intersubband scattering can contribute to the total hole dephasing. 

We compare $\tau_\phi$ extracted in our dots with values measured in the 2DHS. Unfortunately, in GaAs (311) 2DHS, the magnetoconductance around $B = 0 \ \rm T$ originates from a combination of different factors~\cite{Papadakis}, making the extraction of $\tau_\phi$ difficult. Nevertheless, measurements of $\tau_\phi$ have been reported for p-type (100) GaAs and InGaAs quantum wells~\cite{Pedersen}. Extracted values of $\tau_\phi$ in these 2DHSs are consistent with $\tau_\phi$ measured in our quantum dots. This indicates that the small value of $\tau_\phi$ observed in our samples is not determined by the confinement and is likely related to scattering mechanisms in the 2DHS. 

In conclusion, we performed magnetotransport measurements in holes confined to GaAs open quantum dots. We observe clear evidence of coherent transport (MCFs and WAL) inside the dots when $T$ is lowered below 500 mK. We analyse these coherent effects using the RMT developed for spin-$\frac{1}{2}$ carriers. We show that the quantum dots are in a strong spin-orbit coupling regime and extract a hole $\tau_\phi$. While the $T$-depencence of $\tau_\phi$ is qualitatively similar to that reported for electrons,  holes are found to be subject to a stronger dephasing than electrons.  

\begin{acknowledgments}
The authors are indebted to R. Winkler for useful discussions. The work was supported by the DOE and the NSF, the von Humboldt Foundation, "Actions de Recherches Concert\'ees (ARC) - Communaut\'e fran\c{c}aise de Belgique", and by the Belgian Science Policy through the Interuniversity Attraction Pole Program PAI (P5/1/1). S.F. acknowledges financial support from the FRIA and S.M. from the FNRS.
\end{acknowledgments}


\begin{thebibliography}{10}

\bibitem{Spintronic} For reviews, see e.g., G.A. Prinz, Physics Today {\bf 48}, 58 (1995); S.A. Wolf, D.D. Awschalom, R.A. Buhrman, J.M. Daughton, S. von Molnar, M.L. Roukes, A.Y. Chtechelkanova, and D.M. Treger, Science {\bf 294}, 1488 (2001); I. Zutic, J. Fabian and S. Das Sarma, Reviews of Modern Physics {\bf 76}, 323 (2004). 

\bibitem{Bird} For a review, see J.J. Lin and J.P. Bird, J. Phys.: Condens. Matter {\bf 14}, R501 (2002).

\bibitem{dephasingGaAs} R.M. Clarke, I.H. Chan, C.M. Marcus, C.I. Duruoz, J.S. Harris, Jr., K. Campman and A.C. Gossard, Phys. Rev. B {\bf 52}, 2656 (1995); J.P. Bird, K. Ishibashi, D.K. Ferry, Y. Ochiai, Y. Aoyagi and T. Sugano, Phys. Rev. B {\bf 51}, 18037 (1995). 

\bibitem{dephasingGaAs2} A.G. Huibers, M. Switkes, C.M. Marcus, K. Campman and A.C. Gossard, Phys. Rev. Lett. {\bf 81}, 200 (1998).

\bibitem{dephasingInGaAs} B. Hackens, F. Delfosse, S. Faniel, C. Gustin, H. Boutry, X. Wallart, S. Bollaert, A. Cappy and V. Bayot, Phys. Rev. B {\bf 66}, 241305(R) (2002). 

\bibitem{BenPRL} B. Hackens, S. Faniel, C. Gustin, X. Wallart, S. Bollaert, A. Cappy and V. Bayot, Phys. Rev. Lett. {\bf 94}, 146802 (2005). 

\bibitem{Holenano} J.J. Heremans, M.B. Santos and M. Shayegan, Appl. Phys. Lett. {\bf61}, 1652 (1992); I. Zailer, J.E.F. Frost, C.J.B. Ford, M. Pepper, M.Y. Simmons, D.A. Ritchie, J.T. Nicholls and G.A.C. Jones, Phys. Rev. B {\bf 49}, 5101 (1994); J.P. Lu, M. Shayegan, L. Wissinger, U. R\"ossler and R. Winkler, Phys. Rev. B {\bf60}, 13776 (1999); J.B. Yau, E.P. De Poortere and M. Shayegan, Phys. Rev. Lett. {\bf 88}, 146801 (2002); L.P. Rokhinson, V. Larkina, Y.B. Lyanda-Geller, L.N. Pfeiffer and K.W. West, Phys. Rev. Lett. {\bf 93}, 146601 (2004); B. Grbic, R. Leturcq, K. Ensslin, D. Reuter and A.D. Wieck, Appl. Phys. Lett. {\bf 87}, 232108 (2005). 

\bibitem{Loss} D. V. Bulaev and D. Loss, Phys. Rev. Lett. {\bf95}, 076805 (2005). 

\bibitem{Eisenstein} J.P. Eisenstein, H.L. Stormer, V. Narayanamurti, A.C. Gossard and W. Wiegmann, Phys. Rev. Lett. {\bf 53}, 2579 (1984).

\bibitem{Lu} J.P. Lu, J.B. Yau, S.P. Shukla, M. Shayegan, L. Wissinger, U. Rossler and R. Winkler, Phys. Rev. Lett. {\bf 81}, 1282 (1998); S.J. Papadakis, E.P. De Poortere, H.C. Manoharan, M. Shayegan and R. Winkler, Science {\bf283}, 2056 (1999). 

\bibitem{Winkler} R. Winkler, {\it Spin-orbit Coupling Effects in Two-Dimensional Electron and Hole Systems} (Springer, Heidelberg, 2003).

\bibitem{footnote3} GaAs $(311)A$ 2DHSs exhibit a mobility anisotropy [see J.J. Heremans, M.B. Santos, K. Hirakawa and M. Shayegan, J. Appl. Phys. {\bf 76}, 1980 (1994)]. The mobility quoted here was measured along the [$\bar{2}33$] high-mobility direction. 

\bibitem{footnote2} The dots' areas were deduced based on the lithographic dimensions and by taking into account the depletion regions, which were estimated from the effective width of the quantum point contacts, given by the number of modes entering the cavity.

\bibitem{PRLWinkler} R. Winkler, S.J. Papadakis, E.P. De Poortere and M. Shayegan, Phys. Rev. Lett. {\bf84}, 713 (2000).

\bibitem{VanHouten} H. van Houten, C.W.J. Beenakker, P.H.M. van Loosdrecht, T.J. Thornton, H. Ahmed, M. Pepper, C.T. Foxon and J.J. Harris, Phys. Rev. B {\bf37}, R8534 (1988). 

\bibitem{Antiloc} A.M. Chang, H.U. Baranger, L.N. Pfeiffer and K.W. West, Phys. Rev. Lett. {\bf 73}, 2111 (1994). 

\bibitem{zumbuhl} D.M. Zumb\"uhl, J.B. Miller, C.M. Marcus, K. Campman and A.C. Gossard, Phys. Rev. Lett. {\bf 89}, 276803 (2002). 

\bibitem{HoleWAL2D} N.S. Averkiev, L.E. Golub and G.E. Pikus, Solid State Commun. {\bf 107}, 757 (1998).

\bibitem{RMT1} P.W. Brouwer, J.N.H.J. Cremers and B.I. Halperin, Phys. Rev. B {\bf 65}, 081302(R) (2002).  

\bibitem{RMT2} Jan-Hein Cremers, P.W. Brouwer, V.I. Fal'ko, Phys. Rev. B {\bf 68}, 125329 (2003).

\bibitem{footnote4} The determination of the cut-off frequencies $f_c$ was performed as in Ref.~\cite{dephasingInGaAs}. We have $f_c=12$ and $17 \ \rm T^{-1}$ for D1 at $p = 2.3 \times 10^{15} \ \rm m^{-2}$ and $p = 1.7 \times 10^{15} \ \rm m^{-2}$, respectively, and $f_c=37 \ \rm T^{-1}$ for D2 at $p = 2.3 \times 10^{15} \ \rm m^{-2}$. 

\bibitem{Bird1} J.P. Bird, A.P. Micolich, H. Linke, D.K. Ferry, R. Akis, Y. Ochiai, Y. Aoyagi and T. Sugano, J. Phys.: Condens. Matter {\bf 10}, L55 (1998); M. Elhassan, J.P. Bird, R. Akis, D. Ferry, T. Ida and K. Ishibashi, J. Phys.: Condens. Matter {\bf 17}, L351 (2005).

\bibitem{tauphi2D} K.K. Choi, D.C. Tsui and K. Alavi, Phys. Rev. B {\bf 36}, 7751 (1987); B.L. Altshuler, A.G. Aronov and D.E. Khmelnitsky, J. Phys. C {\bf 15} 7367 (1982). 

\bibitem{Papadakis} S.J. Papadakis, E.P. De Poortere, H.C. Manoharan, J.B. Yau, M. Shayegan and S.A. Lyon, Phys. Rev. B {\bf65}, 245312 (2002).

\bibitem{Pedersen} S. Pedersen, C.B. S{\o}rensen, A. Kristensen, P.E. Lindelof, L.E. Golub and N.S. Averkiev, Phys. Rev. B {\bf60}, 4880 (1999); G.M. Minkov, A.A. Sherstobitov, A.V. Germanenko, O.E. Rut, V.A. Larionova and B.N. Zvonkov, Phys. Rev. B {\bf71}, 165312 (2005). 


\end{thebibliography}
\end{document}